# PHYSICAL UNCLONABLE FUNCTION (PUF) BASED RANDOM NUMBER GENERATOR

Ali Sadr, Mostafa Zolfaghari-Nejad

Department of Electrical Engineering, Iran University of Science and Technology

Tehran, Iran

`Sadr@iust.ac.ir, Zolfaghari@ee.iust.ac.ir`

## ABSTRACT

*Physical Unclonable Functions (PUFs) are widely used to generate random Numbers. In this paper we propose a new architecture in which an Arbiter Based PUF has been employed as a nonlinear function in Nonlinear Feedback Shift Register (NFSR) to generate true random numbers. The rate of producing the output bit streams is 10 million bits per second. The proposed RNG is able to pass all NIST tests and the entropy of the output stream is 7.999837 bits per byte. The proposed circuit has very low resource usage of 193 Slices that makes it suitable for lightweight applications.*

## KEYWORDS

*Physical Unclonable Function (PUF), Random Number Generator (RNG), Arbiter based PUF*

## 1. INTRODUCTION

Random sequences are used extensively in cryptographic systems and their applications. Random number generator produces a sequence of zero and one bits. The security of many cryptographic systems depends heavily on generation of unpredictable bit sequences. Random number generators are categorized into Pseudo Random Number Generators and True Random Number Generators.

Pseudo RNG generates a sequence based on a mathematical algorithm by using one-way functions. The security of pseudo random sequence depends on the complexity of its algorithms and functions. Unlike pseudo random sequences, true random sequences are totally unpredictable.

One common circuit to generate pseudo random sequences is linear feedback shift register (LFSR). A linear feedback shift register (LFSR) is a shift register whose input bit is a linear function of its previous state. By connecting a non-linear function of the previous state of shift register to its input, a non-linear feedback shift register (NFSR) will be achieved.

PUF works on the basis of intrinsic differences which exist in manufacture of the electronic parts and equipments. Output of PUF is indeed a function of all these differences. Since the function is very complicated, the produced stream by PUF seems to be completely randomized.

The main purpose of this paper is to design and implement a new PUF-based architecture to generate a sequence of random numbers.

This paper is organized as follows. In section 2 the PUF function and its characteristics are discussed. Section 3 presents a brief review of the related literature. The proposed novel architecture of PUF-based RNG is presented in section 4. Experimental results are given in section 5. Finally, the conclusions are remarked in section 6.

## 2. PUF

A Physical Unclonable Function (PUF) is a function that maps challenges to responses, that is embodied by a physical device, and that verifies the following properties:

1. Easy to evaluate: The physical device is easily capable of evaluating the function in a short amount of time.

2. Hard to characterize: From a polynomial number of plausible physical measurements (in particular, determination of chosen challenge-response pairs), an attacker who no longer has the device, and who can only use a polynomial amount of resources (time, matter, etc...) can only extract a negligible amount of information about the response to a randomly chosen challenge [1].

Most of the PUF designs are based on delay variation of logic and interconnect. Two of the most commonly used PUFs are Arbiter-PUFs [2] and Ring-Oscillator PUFs [3].

Figure 1 shows the block diagram of an arbiter-based PUF.

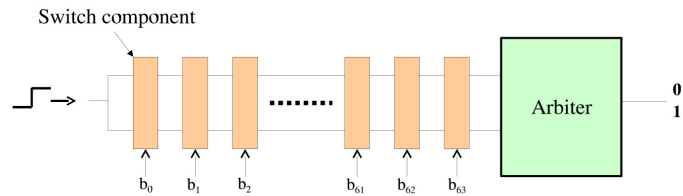

Figure 1. Arbiter based PUF circuit (basic Arbiter scheme). [2]

The main components of this block diagram are switch components and arbiter. Figure 2 shows these components and Figure 3 shows the details of the switch component.

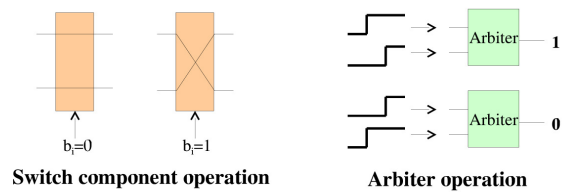

Figure 2. Components of Arbiter based PUF circuit [2]

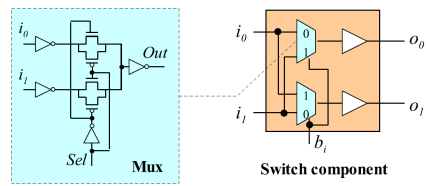

Figure 3. Implementation of a switch component . [2]

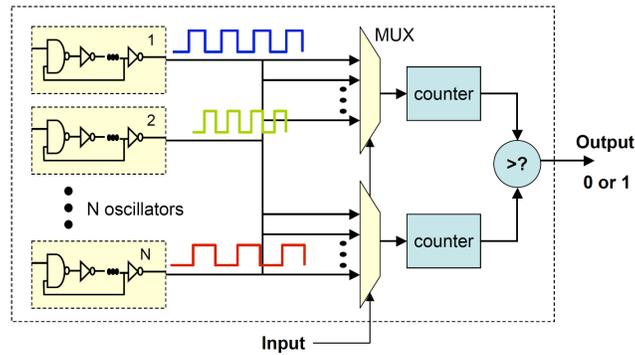

Figure 4. Ring Oscillator Based PUF circuit. [3]

## 3. RELATED WORKS

Ring Oscillator PUFs are used as source of randomness to build TRNGs [4] [5]. Sunar and his co-workers in their research in year 2007 used many distinct ring oscillators arranged on a chip design [6]. Figure 5 shows

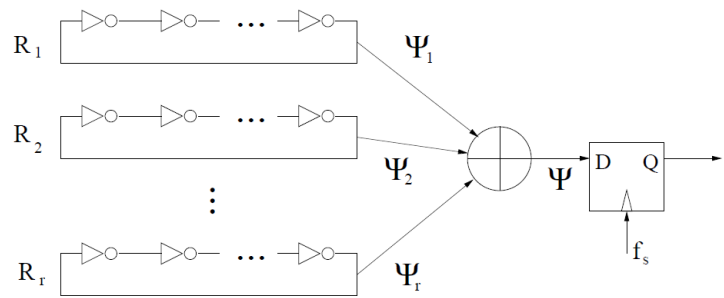

Figure 5. Ring Oscillator Based PUF circuit. [6]

Each ring oscillator is composed of some inverters. The outputs of these ring oscillators are fed to an XOR tree and then sampled at a regular clock frequency.

A true random number generator that exploits metastablity and thermal noise is proposed by Ranasinghe et al [7]. Ayat and his co-workers present a novel architecture which has solved both of security and stability problems [8].

## 4. PROPOSED DESIGN

As mentioned in the introduction, Linear Feedback Shift Register is one of the most widely used methods to generate pseudo random numbers. Since the function located in feedback path of Linear Feedback Shift Register is deterministic, the output sequence will be predictable. Main idea of the proposed True Random Number Generator is to use Physical Unclonable Function as a nonlinear feedback of NFSR. The block diagram of the proposed design is shown in Figure 6.

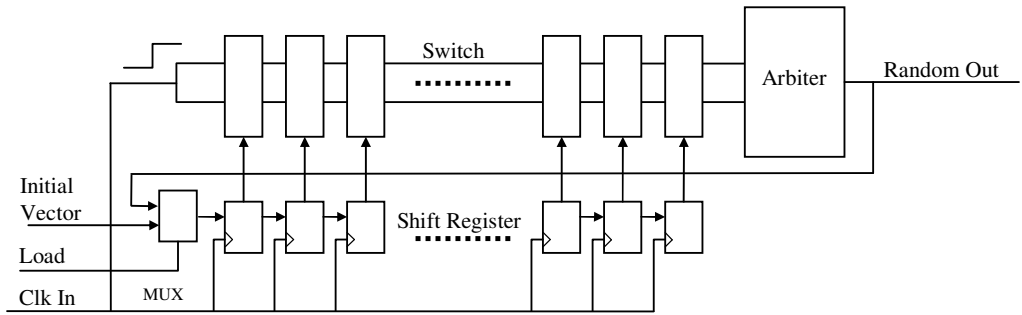

Figure 6. The Block Diagram of Main Idea of Proposed RNG

As observed in the block diagram, output bits of a shift register are presented in parallel as a challenge input to an Arbiter PUF. The output of the PUF depends on input challenge and could be zero or one. As the output of the PUF is a very complicated function of delay parameters of the gates and paths, there will be an unpredictable output.

The main disadvantage of the above diagram is the stability problem of the circuit. In order to increase the stability of the circuit, the comparison has been done by two arbiters. The above arbiter has extra delay on first path and the below arbiter has extra delay on second path.

In first arbiter, if the rising edge of first input comes earlier than the rising edge of second input, the output becomes '1' and in second arbiter, the output becomes '0' if the rising edge of second input comes earlier. The Random output is valid if the outputs of both arbiters have same values; otherwise the output is not valid.

The final architecture is shown in figure 7. In this architecture, the output is valid when the delay difference between two paths is greater than a threshold. By this technique, the effect of those challenges that have a marginal delay difference, will be omitted.

For this architecture, an LFSR with the length of 128 flip-flops has been used. The primitive update function of the LFSR is chosen as equation (1).

$$f(x) = x^{128} + x^{126} + x^{101} + x^{99} + 1 \qquad (1)$$

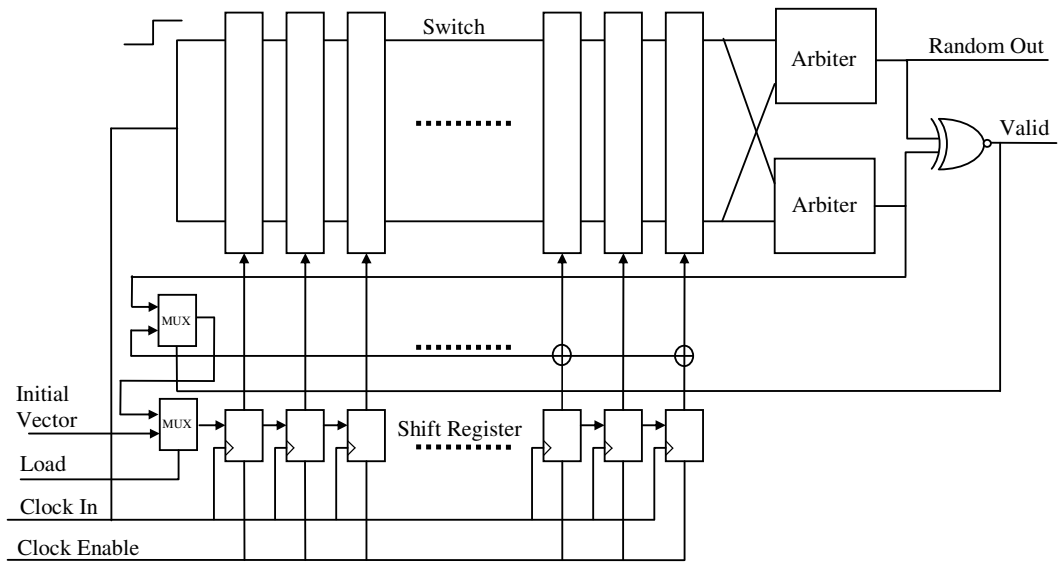

Figure 7. The block diagram of the improved RNG

## 5. EXPERIMENTAL RESULTS

The design is implemented using virtex-4 Xilinx FPGA (XC4VSX55). The Xilinx XC4VSX55 Virtex-4 FPGA has about 24,000 slices. Our 128-stage PUF uses less than 1% of such slices. The clock frequency of the output is 24MHz. A USB port has been used to transfer the output to the connected PC.

Table 1 shows the utilization of resources implementing the proposed design on a Xilinx Virtex 4 device (XC4VSX55FFG1148 -1).

Table 1. Resource utilization of proposed design

| Logic Utilization | Used | Available | Utilization |
|---|---|---|---|
| Number of Slice Flip Flops | 129 | 49,152 | 1% |
| Number of 4 input LUTs | 257 | 49,152 | 1% |
| Number of occupied Slices | 193 | 24,576 | 1% |
|     Number of Slices containing only related logic | 193 | 193 | 100% |
|     Number of Slices containing unrelated logic | 0 | 193 | 0% |
| Total Number of 4 input LUTs | 257 | 49,152 | 1% |
| Number of bonded IOBs | 5 | 640 | 1% |
| Number of BUFG/BUFGCTRLs | 1 | 32 | 3% |
|     Number used as BUFGs | 1 | | |
| Average Fan-out of Non-Clock Nets | 2.33 | | |

Advanced Computing: An International Journal ( ACIJ ), Vol.3, No.2, March 2012In order to evaluate the randomness of the proposed method's output, the test suite of NIST were utilized. For this purpose, an 80 million bit stream was generated in the output.

The NIST Statistical Test Suite includes a set of statistical experiments which attempt to identify the stream of binary numbers which do not behave in a truly random manner. To do this, these tests derive the probability of generating the given sequence by a truly random number generator. A more detailed discussion of the tests and their algorithms can be found in NIST Special Publication [9]. The statistical test suite employed (NIST) consists of sixteen test.

NIST test results are summarized in the table below:

Table 2. NIST Test Results

| Number | Statistical Test | Result |
|---|---|---|
| 1 | Frequency | 100% |
| 2 | Block Frequency | 97% |
| 3,4 | Cumulative Sums | 100% |
| 5 | Runs | 100% |
| 6 | Longest Run | 100% |
| 7 | Rank | 96% |
| 8 | FFT | 96% |
| 10-156 | Non Overlapping Template | 80/80: 50% |
| 157 | Overlapping Template | 100% |
| 158 | Universal | 97% |
| 159 | Approximate Entropy | 97% |
| 160-167 | Random Excursions | 58/58: 37% |
| 168-185 | Random Excursions Variant | 58/58: 87% |
| 186, 187 | Serial | 97%, 96% |
| 188 | Linear Complexity | 98% |

As it is shown in Table 2, all statistical tests are passed.

ENTROPY test results are summarized in the Table II.

Table 3. Entropy Test Results

| Number | Description | Result |
|---|---|---|
| 1 | Entropy | 7.999837 bits per byte |
| 2 | Optimum compression reduction | 0 percent |
| 3 | Chi square distribution | For 9977856 samples is 2252.42, and randomly would exceed this value less than 3.27 percent of the times. |
| 4 | Arithmetic mean value of data bytes | 127.4205 |
| 5 | Monte Carlo value for Pi | 3.144305149 |
| 6 | Serial correlation coefficient | 0.001116 |

As it is shown in table 2, the entropy of the output stream is 7.999837 bits per byte which is so close to the expected value of 8 bits per byte. Also, the optimum compression reduction cannot compress the output stream. The estimated arithmetic mean value of data bytes and the Monte

144

Carlo value of Pi are 127.4205 and 3.144305149 respectively and the expected values of these parameters are 127.5 and 3.1415.

## 5. CONCLUSIONS

In this paper, we propose a true random number generator based on Arbiter PUFs. This architecture increases the stability of the output sequence using two arbiters with a little difference in delay paths. In order to test the randomness of the output stream, the proposed TRNG is implemented on Virtex-4 FPGA and then NIST test suite is applied. The results show that all tests are passed and the output stream is completely random.